\documentclass[prd,nofootinbib,twocolumn,showpacs]{revtex4}  
\usepackage{mathrsfs,graphicx,bm,verbatim}

\usepackage[normalem]{ulem} %to cross out words with \sout

\def\ba{{\bf a}}
\def\bx{{\bf x}}
\def\bn{{\bf n}}

\def\uc{\underline{c}}
\def\us{\underline{s}}
\def\ub{\underline{b}}
\def\urho{\underline{\rho}}
\def\uomega{\underline{\omega}}
\def\ulambda{\underline{\lambda}}
\def\uphi{\underline{\phi}}
     
\begin{document}
\title{Noncommutativity Approach to Supersymmetry on the Lattice:\\
SUSY Quantum Mechanics and an Inconsistency}

\author{Falk Bruckmann}
\affiliation{Instituut-Lorentz for Theoretical Physics,
Leiden University  \\ P.O. Box 9506, NL-2300 RA Leiden, The Netherlands}
                               
\author{Mark de Kok}
\affiliation{Instituut-Lorentz for Theoretical Physics,
Leiden University \\ P.O. Box 9506, NL-2300 RA Leiden, The Netherlands}

%\pacs{}

%\date{\today}
          
\begin{abstract}
It is argued that the noncommutativity approach to fully
supersymmetric field theories on the lattice
suffers from an inconsistency.
Supersymmetric quantum mechanics is worked out in this formalism and the
inconsistency is shown both in general and explicitly for that system, as well as for the Abelian super BF model.
%The inconsistency is shown alsong the lines of 
%supersymmetric quantum mechanics.

\end{abstract}

\maketitle

\section{Introduction}

In supersymmetric theories on the lattice
it is very useful to maintain exact SUSY invariance\footnote{
Just like gauge invariance is kept in lattice gauge theories.},
at least for some of the supersymmetry transformations
(for a review see e.g.\ \cite{Catterall,Giedt1}),
in order to reduce fine-tuning in the continuum limit.
A well-known obstacle against this comes from the failure of the Leibniz
rule (the product rule of differentiation) for the difference operator on the lattice. 
The latter also spoils the Leibniz rule for the supercharges which is extensively
used to write down supersymmetric actions.
%Because of this failure, it is problematic to define supercharges in a superspace representation 
%on the lattice 
%that do obey the Leibniz rule. But in showing supersymmetry invariance of lattice actions
%extensive use of the Leibniz rule is made. Not having such a rule on the lattice therefore makes
%it problematic to define supersymmetric lattice actions. 

D'Adda et al.\ \cite{Dadda,Dadda2,Dadda3} proposed an 
approach using noncommutativity
between the bosonic and fermionic coordinates in a superspace representation.
In this way the Leibniz rule for the supersymmetry transformations is restored
and it is in principle possible to define fully supersymmetric lattice actions in a
straightforward way, merely copying what is done in the continuum. This method
works for two supersymmetric algebras, namely twisted $N=D=2$ and $N=D=4$.

In this paper we will show that the approach can also be used to define
supersymmetric quantum mechanics (SUSYQM, treated as an $N=2$  
one dimensional supersymmetric field theory) on the lattice. With its
minimal field content and its dependency on only one coordinate this system is
the simplest toy model to test nonperturbative features of supersymmetry.
SUSYQM on the lattice has been investigated in \cite{Giedt,CG,Catterall2} keeping half of the
supersymmetries, whereas our approach aims at preserving the full 
supersymmetry nonperturbatively\footnote{A modification of the supersymmetry transformations 
yielding an exact lattice  supersymmetry has been written down in 
\cite{Golterman, Feo}, but to be practical needs to be expanded in powers of the coupling constant.}.

However, in the second part of the paper we will point to
an inconsistency in the whole approach.
Due to the noncommutativity, the supersymmetry transformation of a given
polynomial of fields (e.g.\ in the action) depends on the order of
the fields in it.
Since such a permutation does not change the term (as the fields still
(anti)commute), the definition of the supersymmetry transformation and hence the
question whether a given expression is supersymmetry invariant is not unique.

\section{Supersymmetric Quantum Mechanics}

\subsection{Brief Review in the Continuum}

The algebra of SUSYQM \cite{Cooper1}
can be rewritten in a Majorana representation as
\begin{equation}
\begin{array}{c}
\left\{Q_1,Q_1\right\} = \left\{Q_2,Q_2\right\} = 2H,\\[8pt]  
\left\{Q_1,Q_2\right\}= 0,\quad [H,Q_i] = 0\,\quad \mbox{for }i=1,2,
\end{array}
\end{equation}
%\begin{eqnarray}{c}
%\left\{Q_1,Q_1\right\} = \left\{Q_2,Q_2\right\} = 2H,\nonumber\\
%\left\{Q_1,Q_2\right\}= 0,\quad [H,Q_i] = 0\,\quad \mbox{for }i=1,2,
%\end{eqnarray}
where $Q_1$ and $Q_2$ are two real supercharges and $H$ is the Hamiltonian. 
A superspace representation of this algebra is given by
\begin{equation}
H = \partial_t,\quad
Q_1 = \partial_{\theta^1} + \theta^1 \partial_t,\quad
Q_2 = \partial_{\theta^2} + \theta^2 \partial_t,
\end{equation}
with Majorana Grassmann coordinates $\theta^i$ fulfilling %satisfying the algebra
\begin{equation}
\left\{\theta^i,\theta^j \right\} = 0,\quad
\left\{\partial_{\theta^i},\theta^j \right\} = \delta_{ij},\quad
[t,\theta^i] = 0.
\end{equation}
Having numerical simulations on the lattice in mind,
we have moved to a Euclidean description.

A Hermitian superfield
\begin{equation}
\Phi(t,\theta^1,\theta^2) =
\phi(t) +
i\theta^1 \psi_1(t) +
i\theta^2 \psi_2(t) +
i\theta^2\theta^1 D(t)
\label{eq_superfield_cont}
\end{equation}
contains two real bosonic fields, $\phi$ and $D$, and two Majorana fermions $\psi_{1,2}$.
The action of supersymmetric quantum mechanics is given by
\begin{equation}
S = \int dt d\theta^2 d\theta^1 [\frac{1}{2}D_2\Phi D_1\Phi + i F(\Phi)],
\label{eq_action_cont_super}
\end{equation}
where $F(\Phi)$ is the superpotential. The superderivatives $D_{1,2}$ are
defined as 
\begin{equation}
D_1 = \partial_{\theta^1} -\theta^1\partial_t,\quad D_2
= \partial_{\theta^2} -\theta^2\partial_t.
\end{equation}
They satisfy the following algebra
\begin{equation}
\left\{D_1,D_1\right\} = \left\{D_2,D_2\right\} = -2\partial_t,\quad
\left\{D_1,D_2\right\}= 0,
\end{equation}
and anticommute with the supercharges.

We take the superpotential to be
$F(\Phi)=\frac{1}{2}m\Phi^2 + \frac{1}{4}g\Phi^4$, the choice
$\frac{1}{3}g\Phi^3$ for the interaction term works fully analogously. 
After integrating out the Grassmann variables the action reads
%\begin{equation}
%\begin{array}{c}
%S = \int dt \,( \frac{1}{2}(\partial_t\phi)^2 -
%\frac{1}{2}D^2 -\frac{1}{2}(\psi_1\partial_t\psi_1 +\psi_2\partial_t\psi_2)\\[10pt] 
%-i(m+3g\phi^2) \psi_1\psi_2 - D(m\phi+g\phi^3)).
%\label{action+D}
%\end{array}
%\end{equation}
\begin{eqnarray}
S = \int dt \,[ \frac{1}{2}(\partial_t\phi)^2 -
\frac{1}{2}D^2 -\frac{1}{2}(\psi_1\partial_t\psi_1 +\psi_2\partial_t\psi_2)\nonumber\\
-i(m+3g\phi^2) \psi_1\psi_2 - D(m\phi+g\phi^3)].
\label{action+D}
\end{eqnarray}
The field $D$ is nondynamical and using its equation of motion
$D=-m\phi-g\phi^3$ one arrives at the on-shell action
%\begin{equation}
%\begin{array}{c}
%S = \int dt \,( \frac{1}{2}(\partial_t\phi)^2 +\frac{m^2}{2}\phi^2 -  
%\frac{1}{2}(\psi_1\partial_t\psi_1  + \psi_2\partial_t\psi_2)\\[10pt] 
%- i(m+3g\phi^2) \psi_1\psi_2  + mg\phi^4 + \frac{g^2}{2}\phi^6).
%\end{array}
%\end{equation}
\begin{eqnarray}
S = \int dt \,[ \frac{1}{2}(\partial_t\phi)^2 +\frac{m^2}{2}\phi^2 -  
\frac{1}{2}(\psi_1\partial_t\psi_1  + \psi_2\partial_t\psi_2)\nonumber\\ 
- i(m+3g\phi^2) \psi_1\psi_2  + mg\phi^4 + \frac{g^2}{2}\phi^6].
\end{eqnarray}

By letting the supercharges act on the superfield $\Phi$
it is a straightforward exercise to find
the supersymmetry variations $\delta_1 = \epsilon^1 Q_1$
and $\delta_2 = \epsilon^2 Q_2$ of
the component fields, see Table \ref{table_SQMtrans}.
\begin{table}[h]
 \begin{tabular}{c|c|c}
%\hline
 $\Phi$ & $\delta_1\Phi$ & $\delta_2\Phi$  \\[6pt]
  \hline
  $\phi$ & $i\epsilon^1\psi_1$ & $i\epsilon^2\psi_2$ \\[5pt]
 $\psi_1$ & $i\epsilon^1\partial_t\phi $  & $-\epsilon^2 D$ \\[5pt]
  $\psi_2$ &  $\epsilon^1 D$ & $i\epsilon^2\partial_t\phi $\\[5pt]
  $D$ & $-\epsilon^1\partial_t\psi_2$ &  $\epsilon^2\partial_t \psi_1$ \\[5pt]
%\hline
 \end{tabular}
\caption{Supersymmetry transformations in the continuum.
The same set of variations is valid for the (noncommutative) SUSY
on the lattice replacing $\partial_t$ by $\Delta_\pm$
according to (\ref{eq_charges_lattice}).}
\label{table_SQMtrans}
\end{table}
Omitting the $\epsilon$ parameters, the supersymmetry transformations
of the component fields are denoted by $s_i$, e.g.\ $s_1\phi = i\psi_1$.
Since the supercharges obey the Leibniz rule, it follows that the latter
also holds for the supersymmetry transformations $s_i$:
\begin{equation}
s_i\left[f_1f_2\right]= \left[s_i f_1\right]f_2 + 
                    (-1)^{|f_1|}f_1[s_i f_2],
\label{prodrulefors_i}
\end{equation}
where $|f|$ is $0$ for bosonic $f$ and $1$ for fermionic $f$.

Letting the supersymmetry transformations $s_1$ and $s_2$
act on the action (\ref{action+D}), the Leibniz rule is used to show
that the Lagrangian is supersymmetry invariant (up to a total derivative).

%%%%%%%%%%%%%%%%%%%%%%%%%%%%%%%%%%%%%%%%%%%%%%%%%%%%%%%%%%%%%%%%%%%%%%%%%

\subsection{Definition on the Lattice}

In this section the noncommutativity approach to SUSY on the lattice
\cite{Dadda,Dadda2,Dadda3} is introduced by means of SUSYQM.
Periodic boundary conditions will be assumed for all fields.

Naturally, the derivative $\partial_t$ is replaced on the lattice by the 
forward/backward difference operator 
$\Delta_{\pm}$
\begin{equation}
\Delta_{\pm}f(t) = \pm\frac{1}{2n}\left(f(t\pm 2n) - f(t) \right),
\label{eq_difference_operator}
\end{equation}
where $n$ corresponds to the shift of one lattice spacing.
Why the difference is taken over two lattice spacings will become clear later. 
This difference operator does not obey the conventional Leibniz rule,
but rather a `modified Leibniz rule':
\begin{equation}
\begin{array}{ccl}
\Delta_{\pm}\left[f_1(t)f_2(t)\right]
& = & \left[\Delta_{\pm}f_1(t)\right]f_2(t) +\\[8pt]
 &   & f_1(t\pm 2n)\left[\Delta_{\pm}f_2(t)\right].
\end{array}
\label{mlrsqm}
\end{equation}
Taking the lattice supercharges to be
$Q_i = \partial_{\theta^i} + \theta^i\Delta_\pm$ 
it is obvious that they
do not obey the Leibniz rule either
and a naive approach would run into problems defining supersymmetric actions on the lattice.

The main idea of the noncommutativity approach is to introduce a noncommutativity between
the bosonic and Grassmann coordinates of superspace\footnote{
Interestingly, a calculus with $[x_\mu,dx_\nu]\sim dx_\rho$ has been
worked out in \cite{Dimakis,Kanamori} keeping the Leibniz rule for the discrete
derivative.
It seems attractive to identify the supercoordinates $\theta_{1,2}$ with
differentials $dx_{1,2}$ (which are Grassmannians, too), but this introduces
new coordinates $x_{1,2}$.}
%However, it is not clear how to make use of it for SUSYQM, since the latter
%possesses two Grassmannians $\theta_{1,2}$ but only one differential $\d t$.},
in such a way that both supersymmetry 
transformations $\epsilon^1 Q_1$ and $\epsilon^2 Q_2$ obey the normal Leibniz
rule and then to proceed along the lines of continuum SUSY.

When the lattice supercharges act on products of functions similar to Eq.\ (\ref{mlrsqm}),
$\theta^i$ has to be `pulled through' the first function $f_1$ and this can
be used to restore the Leibniz rule for $\theta^i\Delta_{\pm}$ if one demands
\begin{equation}
\theta^{i}f(t) = f(t - a_i)\theta^{i} \Leftrightarrow 
[t,\theta^{i}] = a_i\theta^i\quad \mbox{(no sum)},
\label{eq_ncrel_1}
\end{equation}
with $a_i=\pm 2n$.
However, this new relation implies another noncommutativity 
between $\partial_{\theta^{i}}$ and $t$, with the opposite sign
\begin{equation}
[t,\partial_{\theta^i}] = -a_i\partial_{\theta^{i}} \quad \mbox{(no sum)},
\label{eq_ncrel_2}
\end{equation}
which makes it impossible to obtain the Leibniz rule for both terms in the
supercharges at the same time.

In contrast, the supersymmetry transformations $\epsilon^i Q_i$ (no sum) can
obey the Leibniz rule, provided the same noncommutativity is introduced for
the variation parameters $\epsilon^{i}$
\begin{equation}
[t,\epsilon^{i}] = a_i\epsilon^i \quad \mbox{(no sum)}.
%,\quad[t,\partial_{\epsilon^i}] = -a_i\partial_{\epsilon^{i}}.
\label{eq_ncrel_3}
\end{equation}
Now the Leibniz rule is obvious for the terms $\epsilon^i\partial_{\theta^i}$,
while for the terms $\epsilon^i\theta^i\Delta_{\pm}$ it leads to $2a_i=\pm2n$.
The corresponding Jacobi identities have been checked to give no further constraints.

Altogether, the supercharges
\begin{equation}
Q_1 = \partial_{\theta^1} + \theta^1\Delta_{\pm},\quad
Q_2 = \partial_{\theta^2} + \theta^2\Delta_{\pm},
\label{eq_charges_lattice}
\end{equation}
(with independent signs) fulfill the algebra
\begin{eqnarray}
\left\{Q_1,Q_1\right\} = 2\Delta_\pm,\:
\left\{Q_2,Q_2\right\} = 2\Delta_\pm,\:
\left\{Q_1,Q_2\right\} = 0
\label{eq_algebra_lattice}
\end{eqnarray}
with the same signs as in Eq.\ (\ref{eq_charges_lattice}).
Their supersymmetry transformations $\epsilon^1 Q_1$ and
$\epsilon^2 Q_2$ obey the Leibniz rule,
provided the noncommutativity relations
(\ref{eq_ncrel_1}), (\ref{eq_ncrel_2}) and (\ref{eq_ncrel_3}) hold with
\begin{equation}
a_1=\pm n,\quad a_2=\pm n,
\label{eq_shifts}
\end{equation}
again with the signs as in Eq.\ (\ref{eq_charges_lattice}). The Grassmannians
$\theta^i$ and $\epsilon^i$ still anticommute.

Having a normal Leibniz rule for the supersymmetry transformations,
we proceed to build a supersymmetric lattice action. We take the Hermitian
superfield (\ref{eq_superfield_cont}) like in the continuum with $t$ now
labeling the lattice points.
%For the supersymmetry transformations $s_i$ the noncommutativity of
%$\theta^i$ and $\epsilon^i$ does not matter,
Acting with the supercharges on it, one immediately sees that 
Table \ref{table_SQMtrans} is also valid on the lattice
upon replacing $\partial_t$ by $\Delta_\pm$. 
But the noncommutativity shows up when moving
the $\theta$'s through the fields, e.g.\
\begin{equation}
\begin{array}{ccl}
\Phi & = & \phi(t) - i \psi_1(t-a_1) \theta^1 - i \psi_2(t-a_2) \theta^2  \\[8pt]
&&+ i D(t-a_1-a_2) \theta^2\theta^1,
\label{eq_theta_fields} 
\end{array}
\end{equation}
and especially when bringing the $\theta$'s to the left in the 
computation of the action.

From (\ref{eq_theta_fields}) it can be seen that $t-a_i=t\pm n$
needs to be a lattice point, which is why we
took the difference over two lattice spacings in Eq.\
(\ref{eq_difference_operator}).

In order to construct the kinetic term for this multiplet,
we use the lattice superderivatives
\begin{equation}
D_1 = \partial_{\theta^1} - \theta^1\Delta_{\pm},\quad
D_2 = \partial_{\theta^2} - \theta^2\Delta_{\pm}
\label{eq_superderivatives_lattice}
\end{equation}
with the signs as in Eq.\ (\ref{eq_charges_lattice}) and allow for a relative
shift $\alpha$ in $D_2\Phi D_1\Phi$:
\begin{eqnarray}
&&n\sum_t \int d\theta^2 d\theta^1 \frac{1}{2}D_2\Phi(t-\alpha) D_1\Phi
(t)= \\
&& \frac{n}{2}\sum_t [ \Delta_\pm\phi(t+a_1-\alpha)\Delta_\pm\phi(t)
-D(t+a_2-\alpha)D(t)\nonumber\\
&&-(\psi_1(t)\Delta_\pm\psi_1(t-\alpha)+\psi_2(t+a_1+a_2-\alpha)
\Delta_\pm\psi_2(t))].\nonumber
\end{eqnarray}
%\begin{equation}
%\begin{array}{l}
%n\sum_t \int d\theta^2 d\theta^1 \frac{1}{2}D_2\Phi(t-\alpha) D_1\Phi
%(t)\nonumber\\[10pt]
%= \frac{n}{2}\sum_t (\Delta_\pm\phi(t+a_1-\alpha)\Delta_\pm\phi(t)
%-D(t+a_2-\alpha)D(t)\nonumber\\[10pt]
%-(\psi_1(t)\Delta_\pm\psi_1(t-\alpha)+\psi_2(t+a_1+a_2-\alpha)\Delta_\pm\psi_2(t)))
%\end{array}
%\end{equation}
This action is supersymmetry invariant for any choice of the shift $\alpha$.

At this point we invoke two physical arguments.
When integrating out $D$ later, it is desirable to have an equation of motion
like $D(t)=f[\phi]$ which comes out only if the corresponding term in
the action 
reads $D^2(t)/2$, i.e. if the shift $\alpha$ is fixed to $a_2$ (all other choices
%including the naive one $\alpha=0$
would lead to a nonlocal equation
$D(t+T)+D(t-T)=f[\phi]$ with some $T$).

Moreover, we wish to obtain a strictly positive kinetic term
$(\Delta\phi(t))^2/2$ for the bosonic component. This gives more constraints
depending on the choice of the difference operators in (\ref{eq_charges_lattice}).
%$\Delta_{1,2}=\Delta_\pm$.
By virtue of $\Delta_+ f(t)=\Delta_- f(t+2n)$ and
changes in the summation variable $t$ all solutions to these constraints are
equivalent to taking just the plus signs in
Eq.s (\ref{eq_charges_lattice}), (\ref{eq_algebra_lattice}),
(\ref{eq_shifts}) and (\ref{eq_superderivatives_lattice}).
This leads to the following kinetic part of the action
\begin{eqnarray}
&&\frac{n}{2}\sum_t [(\Delta_+\phi(t))^2-D^2(t)\nonumber\\
&&-(\psi_1(t+n)\Delta_+\psi_1(t)+\psi_2(t+n)\Delta_+\psi_2(t))].
\end{eqnarray}
Observe that the argument $t+n$ of the spinors $\psi_{1,2}$ is right in the middle of
the two lattice points used in the difference operator.

For the mass and interaction terms there seems to be no particular reason to introduce
relative shifts and we find the lattice analogue of the continuum action
(\ref{action+D}) to be 
\begin{eqnarray}
S&=&n\sum_t [ \frac{1}{2}(\Delta_+\phi)^2 - \frac{1}{2}D^2
\nonumber\\
&&-\frac{1}{2}[\psi_1(t+n)\Delta_+\psi_1+\psi_2(t+n)\Delta_+\psi_2]\nonumber\\
&&-i[m+3g\phi^{(2)}][\psi_1(t+n)\psi_2-\psi_2(t+n)\psi_1]/2\nonumber\\
&&-D[m\phi^{(1)}+g\phi^{(3)}]],\label{eq_action_lattice+D}
\end{eqnarray}
where unshifted arguments $(t)$ are not displayed for readability.
We used $\phi^{(a)}$ to abbreviate the lattice terms
that converge to the corresponding powers $\phi^a$ in the continuum limit
\begin{eqnarray}
\!\!\!\!\!\!\phi^{(1)}(t)&\!\!=&\!\![\phi(t+2n)+\phi(t)]/2, \label{phia1}\\
\!\!\!\!\!\!\phi^{(2)}(t)&\!\!=&\!\![[\phi(t+2n)+\phi(t+n)][\phi(t+n)+\phi(t)]\nonumber\\
      &&\!\!+\phi^2(t+2n)+\phi^2(t)]/6,\label{phia2}\\
\!\!\!\!\!\!\phi^{(3)}(t)&\!\!=&\!\![\phi^2(t+2n)+\phi^2(t)][\phi(t+2n)+\phi(t)]/4.\label{phia3}
%\label{phia3}
\end{eqnarray}
While the kinetic part of the lattice action (\ref{eq_action_lattice+D})
only differs by one shift from the naive translation of
the continuum, the complicated structure of the $\phi^{(a)}$'s
is the nontrivial result of the noncommutativity approach and is needed for
supersymmetry invariance to be discussed in Section \ref{Explicitcalculation}. 
However, the arguments of the fields in the $\phi^{(a)}$'s are never shifted over
more than two lattice spacings.

Like in the continuum, the field $D$ can be replaced by
$-m\phi^{(1)}-g\phi^{(3)}$.
We will consider this action in slightly more detail now, in order to compare to
existing SUSYQM lattice actions. With Dirac spinors
$\psi=(\psi_1+i\psi_2)/\sqrt{2}$ it reads
\begin{eqnarray}
S^{\rm on}&=&n\sum_t [\frac{1}{2}(\Delta_+\phi)^2 +
\frac{1}{2}(m\phi^{(1)}+g\phi^{(3)})^2\\
&&-\bar{\psi}(t+n)\Delta_+\psi
-(m+3g\phi^{(2)})\bar{\psi}(t+n)\psi^{(1)}],\nonumber
\end{eqnarray}
where the fermion enters the mass and interaction
term in the form $\psi^{(1)}$ of (\ref{phia1}), too.
Without interactions we write the action in the bilinear form
\begin{eqnarray}
S^{\rm on}_{g=0}=n\sum_{t,t'}[
\frac{1}{2}\phi(t)B_{tt'}\phi(t')+
\bar{\psi}(t)F_{tt'}\psi(t')],
\end{eqnarray}
with the following bosonic and fermionic matrix:
\begin{eqnarray}
B_{tt'}&=&\frac{1}{2}(\frac{1}{n^2}+m^2)\delta_{tt'}\nonumber\\
&&+\frac{1}{4}(-\frac{1}{n^2}+m^2)(\delta_{t+2,t'}+\delta_{t-2,t'}),\\
F_{tt'}&=&\frac{1}{2}(\frac{1}{n}-m)\delta_{t-1,t'}
+\frac{1}{2}(-\frac{1}{n}-m)\delta_{t+1,t'}.
\end{eqnarray}
The supersymmetry of the action expresses itself by the fact that
$B=F^TF$ such that the corresponding determinants in the path integral cancel
each other.

It can straightforwardly be seen that in the noninteracting case our action
amounts to the one given in \cite{Giedt} by Eq.\ (2.18) putting $r/a=-m$, up 
to a relative minus sign for the fermion terms\footnote{
We follow the conventions of \cite{Cooper1}.}.
%{\tt (with $a$ being the lattice spacing in that reference)?}.
However, for the case of a $\Phi^4$-interaction that action gives a diagonal
term coupling the fermions to $\phi^2$, while in our case there will always
be terms like $\phi(.)\phi(.)\bar{\psi}(t\pm n)\psi(t)$.

%%%%%%%%%%%%%%%%%%%%%%%%%%%%%%%%%%%%%%%%%%%%%%%%%%%%%%%%%%%55

\section{The Inconsistency}

\subsection{The General Noncommutativity Approach and its Inconsistency}

In this section we will show that the noncommutativity approach to lattice SUSY suffers
from an inconsistency, independent of the dimensionality of space-time and the
number of supersymmetries.
Before doing so, we first review the general formalism briefly.

In higher dimensional spaces,  like in the case of SUSYQM,
the derivative operator $\partial_{\mu}$ is replaced on the lattice by the 
difference operator 
$\Delta_{\pm\mu}$ defined by
\begin{equation}
\Delta_{\pm\mu}f(\bx) = \pm\frac{1}{2|\bn_{\mu}|}\left(f(\bx\pm 2\bn_{\mu}) - f(\bx) \right),
\end{equation}
where $\bn_{\mu}$ corresponds to the shift of one lattice spacing in the $\mu$-direction.
Each $\Delta_{\pm\mu}$ obeys the modified Leibniz rule (\ref{mlrsqm}).

In a superspace representation 
with coordinates  $\left(x^{\mu}, \theta^A \right)$
the supercharges can generically be written as 
\begin{equation}
Q_A = \partial_{\theta_A} +
\frac{1}{2}f^{\mu}_{AB}\theta^B\Delta_{\pm\mu},
\end{equation}
where we have immediately moved to the lattice using $\Delta_{\pm\mu}$.
The general SUSY algebra reads
\begin{equation}
\left\{Q_A,Q_B \right\} = f^{\mu}_{AB}\Delta_{\pm\mu}.
\end{equation}
In complete analogy to the case of SUSYQM the supercharges will not obey
the Leibniz rule, but the variations $\delta_A=\epsilon^A Q_A$ do,
provided the noncommutativities
\begin{equation}
[\bx,\theta^{A}] = \ba_A\theta^{A}, \quad [\bx,\epsilon^{A}] =
\ba_A\epsilon^{A}\quad \mbox{(no sum)}
\label{eq_nc_general}
\end{equation}
hold. The shift parameters $\ba_A$ (now vectors) ensure the Leibniz rule
for the first term in $Q_A$, while the second one 
imposes the following conditions
\begin{equation}
\ba_A + \ba_B = \pm 2\bn_{\mu} \mbox{ for } f^{\mu}_{AB} \neq 0. 
\label{relations}
\end{equation}
As before, the Jacobi identities are seen to give no further constraints.

The authors of \cite{Dadda}
have found a solution to these relations for the twisted
$N=2$  supersymmetry algebra in two dimensions as
well as for the  twisted $N=4$  supersymmetry algebra in four dimensions.
As explained above, the relations can also be satisfied for
the $N=2$ supersymmetry algebra in one dimension.

At this point one can proceed as in the continuum with building 
a supersymmetric action out of (chiral) superfields.
The superderivatives
\begin{equation}
D_A = \partial_{\theta_A} -
\frac{1}{2}f^{\mu}_{AB}\theta^B\Delta_{\pm\mu}
\end{equation}
obey the algebra
\begin{equation}
\left\{D_A,D_B \right\} = -f^{\mu}_{AB}\Delta_{\pm\mu}, \quad
\left\{D_A,Q_B\right\} = 0.
\label{superderivatives}
\end{equation}

The transformation rules $\epsilon_As_A$ for the component fields 
(like in Table \ref{table_SQMtrans}) can be derived by comparing the
$\theta$-expansions of the superfields $\Phi_i$
of the theory with their supersymmetry 
variations $\epsilon_A Q_A \Phi_i$ or, equivalently, by doing the same with the 
superfields $D_A\Phi_i, D_AD_B\Phi_i,\ldots$ (because  the superderivatives and 
supercharges anticommute).
\\

However, this noncommutativitiy approach is spoiled by an inconsistency.
At the heart of it lies the fact that the supersymmetry 
transformations $s_A$ obey a modified Leibniz rule when acting on a product of
(component) fields.
This can be seen by evaluating the transformations of a product of fields
\begin{eqnarray}
f_1f_2&\rightarrow&(f_1+\epsilon_As_Af_1)(f_2+\epsilon_As_Af_2)\nonumber\\
&=&f_1f_2+[\epsilon_As_Af_1]f_2+f_1[\epsilon_As_Af_2]+O(\epsilon^2)\nonumber\\
&\equiv&f_1f_2+\epsilon_As_A[f_1f_2] +O(\epsilon^2)
\label{preprodrule}
\end{eqnarray}
This is the normal Leibniz rule for the supersymmetry transformations  $\epsilon_As_A$. 
For the transformations $s_A$ we bring
$\epsilon_A$ to the left using (\ref{eq_nc_general}), omit it, and
obtain\footnote{
A specific case of this formula, namely the supersymmetry transformation 
of the product of two particular fields in the lattice super BF model, has been
written down in \cite{Dadda2} as Eq.\ (18).}
\begin{equation}
\begin{array}{l}
s_A\left[f_1(\bx)f_2(\bx)\right]  = \\[8pt] 
\left[s_A f_1(\bx)\right]f_2(\bx)
+ (-1)^{|f_1|}f_1(\bx+\ba_A)\left[s_A f_2(\bx)\right].
\end{array}
\label{prodrule}
\end{equation}

Because the product $f_1f_2$ of component fields is equal 
to $(-1)^{|f_1||f_2|}f_2f_1$, one 
would expect the corresponding supersymmetry transformations to agree, too.
However, because of the different shifts induced by $\epsilon^A$, one gets
\begin{equation}
\begin{array}{l}
s_A\left[(-1)^{|f_1||f_2|}f_2(\bx)f_1(\bx)\right] = \\[8pt]
\left[s_Af_1(\bx)\right]f_2(\bx+\ba_A) +
(-1)^{|f_1|}f_1(\bx)\left[s_Af_2(\bx)\right].
\label{prodrule_again}
\end{array}
\end{equation}
This means that
{\em the supersymmetry transformations do not evaluate to the
same expression when acting on the product $f_1f_2$
and the equivalent expression $(-1)^{|f_1||f_2|}f_2f_1$, respectively.}
Instead, the difference of the two can be written as 
\begin{equation}
\begin{array}{l}
\left[s_Af_1(\bx)\right](f_2(\bx+\ba_A)-f_2(\bx)) \\[8pt]
+(-1)^{|f_1|}(f_1(\bx)-f_1(\bx+\ba_A))\left[s_Af_2(\bx)\right],
\label{eq_difference}
\end{array}
\end{equation}
which is not even a total difference, i.e.\ does not lead to a boundary term 
(in which case the difference would be irrelevant).

Because all shifts are proportional to the lattice spacing, the ambiguity
disappears in the continuum limit, but to keep exact supersymmetry at finite
spacing remains a problem\footnote{
From the corresponding modified Leibniz rule for the difference operator
one might be tempted to conclude that this operator suffers from the
same problem, but its definition $\pm 2n\Delta_\pm[f_1f_2]=[f_1f_2](t\pm
2n)-[f_1f_2](t)$ is invariant under interchanging $f_1$ and $f_2$.}.\\

In the derivation we have taken the natural point of view that the supersymmetry
transformations of all combinations of component fields follow -- via the Leibniz rule
(\ref{preprodrule}) -- from the transformations of the individual fields,
Table \ref{table_SQMtrans}. Having obtained an action
and the supersymmetry transformations of component fields (plus the
noncommutativity of $t$ and $\epsilon$) one should be
allowed to neglect the auxiliary superspace formalism.
The lattice action resulting from this
formalism is a particular
discretization of the continuum action, but the supersymmetry transformations
have been shown to be inconsistent.

Supersymmetry transformations of products of component fields derived from products
of superfields, e.g.\ by $\theta$-expanding $\Phi^2$ and $\delta_A \Phi^2$, should
agree with the ones used so far for consistency.
If {\em other} supersymmetry rules for the products of component fields are obtained in
this way, then an unacceptable memory to superspace is introduced and,
what is more, products of fields are treated like new independent fields.

Also in this approach we are able to derive the inconsistency. Any of the
superfields $\Phi_i,\,D_A\Phi_i,\,D_A D_B\Phi_i,\ldots$ can be written as
\begin{eqnarray}
F=f+\sum_B \theta^B s_B f+O(\theta^2),
\end{eqnarray}
because then $\epsilon^A Q_A F=\epsilon^A s_A f+O(\theta)$ gives the correct
supersymmetry transformation for $f$. This first component $f$ runs over all
component fields in the theory when taking all such 
superfields $F$. Now we consider the
product
\begin{eqnarray}
F_1(\bx)F_2(\bx)&&\!\!\!\!=f_1(\bx)f_2(\bx)+\sum_B\theta^B
\{[s_B f_1(\bx)]f_2(\bx)+\nonumber\\
&&\!\!\!\!(-1)^{|f_1|}f_1(\bx+\ba_B)[s_B f_2(\bx)]\}+O(\theta^2),
\label{eq_sfieldprod_one}
\end{eqnarray}
the variation of which is given by
\begin{eqnarray}
\epsilon^A Q_A [F_1(\bx)F_2(\bx)]&&\!\!\!\!=
\epsilon^A\{[s_A f_1(\bx)]f_2(\bx)+\label{eq_sfieldprod_two}\\
&&\!\!\!\!(-1)^{|f_1|}f_1(\bx+\ba_A)[s_A f_2(\bx)]\}+O(\theta).\nonumber
%\label{eq_sfieldprod_two}
\end{eqnarray}
From the first components of (\ref{eq_sfieldprod_one})
and (\ref{eq_sfieldprod_two}) we read off Eq.\ (\ref{prodrule}) for the
supersymmetry transformation of the product of any two component fields.

In general, the product of two superfields depends on their order 
(because of the noncommutativity (\ref{eq_nc_general})),
but the first component is always the same (up to a sign).
Therefore, looking in the same manner at the variation of 
\begin{eqnarray}
F_2(\bx)F_1(\bx)&&\!\!\!\!=f_2(\bx)f_1(\bx)+\sum_B\theta^B
\{[s_B f_2(\bx)]f_1(\bx)+\nonumber\\
&&\!\!\!\!(-1)^{|f_2|}f_2(\bx+\ba_B)[s_B f_1(\bx)]\}+O(\theta^2)
\end{eqnarray}
and multiplying with the factor $(-1)^{|f_1||f_2|}$ one arrives at
Eq.\ (\ref{prodrule_again}) for the
supersymmetry transformation of the product of two component fields with the
opposite order. Hence, the inconsistency is rederived, because it is the
disagreement between 
(\ref{prodrule}) and (\ref{prodrule_again}).\\

As a consequence, there is {\em
an ambiguity in showing supersymmetry invariance of lattice actions}. 
Indeed, suppose starting with a term $f_1f_2$ in a lattice action $S$ one can 
show supersymmetry invariance of this action, i.e.\ $s_A S = 0$.
The same action will not be supersymmetry invariant when writing this term as
$(-1)^{|f_1||f_2|}f_2f_1$, because now $s_A S$ will be equal to 
a difference term of the form (\ref{eq_difference}),
which does not vanish.
Therefore, the noncommutativity approach is seen to be inconsistent. 

%%%%%%%%%%%%%%%%%%%%%%%%%%%%%%%%%%%%%%%%%%%%%%%%%%%%%%%%

\subsection{Explicit Calculation}
\label{Explicitcalculation}

\subsubsection*{Supersymmetric Quantum Mechanics}

In this section we would like to demonstrate how the inconsistency emerges
when checking the supersymmetry invariance
of the proposed SUSYQM action (\ref{eq_action_lattice+D}).
We do this by explicitly calculating the
supersymmetry variation of the mass terms (which of course have to be invariant on
their own).
For the variation under $\delta_1$ one can group the four terms
proportional to $m$ into two pairs $M$ and $N$ (which, however, get mixed under
$\delta_2$)
\begin{eqnarray}
M&=&D(t)\phi(t)-i\psi_2(t+n)\psi_1(t),\\
N&=&D(t)\phi(t+2n)+i\psi_1(t+n)\psi_2(t).
\end{eqnarray}
Using the Leibniz rule for $\delta_1$ and the supersymmetry variations of Table
\ref{table_SQMtrans} we get
\begin{eqnarray}
\!\!\!\!\!\!\delta_1 M&\!\!=&\!\!-\epsilon^1\Delta_+\psi_2(t)\phi(t)+iD(t)\epsilon^1\psi_1(t)\nonumber\\
&&\!\!-i\epsilon^1D(t+n)\psi_1(t)+\psi_2(t+n)\epsilon^1\Delta_+\phi(t).
\end{eqnarray}
Moving $\epsilon^1$ to the left results in two shifts (and a sign change)
\begin{eqnarray}
\delta_1 M&=&\epsilon^1[-\Delta_+\psi_2(t)\phi(t)
+iD(t+n)\psi_1(t)\nonumber\\
&&-iD(t+n)\psi_1(t)-\psi_2(t+2n)\Delta_+\phi(t)]\nonumber\\
&=& -\epsilon^1\Delta_+[\psi_2(t)\phi(t)].
\end{eqnarray}
Since the total difference vanishes under the sum over $t$, this term is
supersymmetry invariant as it should. The various terms cancel as in
the continuum.

However, if we interchange $D(t)$ and $\phi(t)$, which does not change $M$,
\begin{eqnarray}
\tilde{M}=\phi(t)D(t)-i\psi_2(t+n)\psi_1(t),
\end{eqnarray}
then $\epsilon^1$ induces different shifts leading to
\begin{eqnarray}
\delta_1 \tilde{M}&=&\epsilon^1[i\psi_1(t)D(t) -\phi(t+n)\Delta_+\psi_2(t) \nonumber\\
&&-iD(t+n)\psi_1(t)-\psi_2(t+2n)\Delta_+\phi(t)]\nonumber\\
&\neq & \epsilon^1\cdot\mbox{ total difference}.
\end{eqnarray}
Hence, written in this form, the term turns out not to be supersymmetry
invariant.

Actually, the other mass term $N$ is not invariant as it stands,
but
\begin{eqnarray}
\tilde{N}=\phi(t+2n)D(t)+i\psi_1(t+n)\psi_2(t)
\end{eqnarray}
is.

Since $M=\tilde{M}$ and $N=\tilde{N}$ (the fields are treated
as (anti)commuting functions in a path integral),
the definition of the supersymmetry transformations
of the mass terms is not unique,
nor is the question whether (the sum of) these terms are supersymmetry
invariant.

Of course this problem does not occur for squares
(like $(\Delta_+\phi)^2$ in (\ref{eq_action_lattice+D})).
It can be checked that it neither appears
for the kinetic term of the spinors (the second line in
(\ref{eq_action_lattice+D})).
However, the ambiguity problem persists for the on-shell action 
(which is not just a sum of squares) because of the third line 
in (\ref{eq_action_lattice+D}), which as a fermionic term does not
change when integrating out $D$.
\\

\subsubsection*{The Abelian Super BF Model}

As a second concrete example of the inconsistency,
the Abelian super BF model in two dimensions is investigated. 
This model has been formulated
using twisted $N=2$ supersymmetry in Paragraph 5.1 of \cite{Dadda}
(the conventions of \cite{Dadda} are followed).
In this algebra, there are four different supersymmetry 
transformations, $\us, \tilde{\us}, \us_1, \us_2$, each with its own
shift parameter, respectively $\ba,\tilde{\ba}, \ba_1, \ba_2$. The relations 
(\ref{relations}) can in this case be 
solved and written as
\begin{equation}
\ba+\tilde{\ba}+\ba_1+\ba_2 = 0,\,\, \ba + \ba_1 = \bn_1,\,\, \ba+\ba_2 = \bn_2.
\label{twN=2rel}
\end{equation}

Following the noncommutativity approach
the lattice action for the abelian super BF model is given by
\begin{equation}
\begin{array}{l}
S_{\mathrm{BF}}^{\mathrm{lat}} = 
\sum_\bx\{\uphi(\bx-\tilde{\ba})\epsilon_{\mu\nu}\Delta_{+\mu}\uomega_{\nu}(\bx+\ba_{\nu})\\[10pt] 
+\ub(\bx-\ba)\Delta_{-\mu}\uomega_{\mu}(\bx+\ba_{\mu}) 
-i\bar{\uc}(\bx)\Delta_{+\mu}\Delta_{-\mu}\uc(\bx)\\[10pt] + i\urho(\bx-\ba-\tilde{\ba})\ulambda(\bx-\ba-\tilde{\ba})\}.
\end{array}
\end{equation}

The supersymmetry transformations for $\us$ are given in the table
below\footnote{
In the conventions of \cite{Dadda} the 
transformation $\us$ `carries a shift', e.g.\ $\us\bar{\uc}(\bx) = -i\ub(\bx+\ba)$, 
whereas the transformations $s_i$ in our conventions do not,
e.g.\ $s_1\phi(x) = i\psi_1(x)$, 
Table~\ref{table_SQMtrans}. This difference in convention is due to a 
different way of defining the superfield, compare Eq.~(4.29) or (5.1)
in \cite{Dadda} with our Eq.~\ref{eq_superfield_cont}.}
and $\us$ obeys a Leibniz rule as in (\ref{prodrule}),
with $\ba_A$ taken to be $2\ba$ (cmp.\ Eq.\ (18) in \cite{Dadda2}).
 
%\begin{center}
\begin{table}[h]
 \begin{tabular}{c|c}
%\hline
 $\uphi$ & $\us\uphi$  \\[6pt]
  \hline
  $\ub(\bx+\ba)$ & $0$  \\[5pt]
 $\uphi(\bx+\tilde{\ba})$ & $i\urho(\bx+\ba+\tilde{\ba})$ \\[5pt]
  $\uomega_\nu(\bx+\ba_{\nu})$& $\Delta_{+\nu}\uc(\bx)$ \\[5pt]
  \hline
 $\ulambda(\bx+\ba_1+\ba_2)$ &   $\epsilon_{\rho\sigma}\Delta_{+\rho}\uomega_{\sigma}(\bx+\ba_{\sigma})$ \\[5pt]
   $\bar{\uc}(\bx)$ & $-i\ub(\bx+\ba)$ \\[5pt]
  $\uphi(\bx+\tilde{\ba})$ & $i\urho(\bx+\ba+\tilde{\ba})$ \\[5pt]
  $\urho(\bx+\ba+\tilde{\ba})$ & $0$ \\[5pt]
  $\uc(\bx)$ & $0$   \\[5pt]
%\hline
 \end{tabular}
\caption{The $N=2$ twisted supersymmetry transformation $\us$ for the abelian
super BF model, copied from \protect\cite{Dadda}.
The fields above the line are bosonic, the ones below fermionic.}
% \end{tabular}
\label{ABFtrans}
\end{table}

%\end{center}
Applying $\us$ to the action it follows that
\begin{equation}
\begin{array}{l}
\us S^{\mathrm{lat}}_{\mathrm{BF}} = 
\sum_\bx \left[i\urho(\bx+\ba-\tilde{\ba})\epsilon_{\mu\nu}\Delta_{+\mu}\uomega_{\nu}(\bx+\ba_{\nu}) \right. \\[10pt]
\,\,\,\,\,\,+\uphi(\bx-\tilde{\ba}+2\ba)\epsilon_{\mu\nu}\Delta_{+\mu}\Delta_{+\nu}\uc(\bx) \\[10pt]
\,\,\,\,\,\, +\ub(\bx+\ba)\Delta_{+\mu}\Delta_{-\mu}\uc(\bx) -\ub(\bx+\ba)\Delta_{+\mu}\Delta_{-\mu}\uc(\bx) \\[10pt]
\,\,\,\,\,\, \left. -i\urho(\bx+\ba-\tilde{\ba})\epsilon_{\rho\sigma}\Delta_{+\rho}\uomega_{\sigma}(\bx+ \ba_{\sigma}-\ba_1-\ba_2-\ba-\tilde{\ba}) \right] \\[10pt]
=  0,
\end{array}
\end{equation}
where identities for the shift parameters
and $\epsilon_{\mu\nu}\Delta_{+\mu}\Delta_{+\nu} = 0$ have been used. 
In other words, the action seems to be invariant under
the supersymmetry transformation $\us$.

Up to this point we agree with the authors of \cite{Dadda}.
However, interchanging the first two fields in the action, i.e.\ writing the action as
\begin{equation}
S_{BF}^{\mathrm{lat}} = \sum_\bx \left[ \epsilon_{\mu\nu}\Delta_{+\mu}\uomega_{\nu}(\bx+\ba_{\nu})\uphi(\bx-\tilde{\ba}) + \mbox{ rest unchanged} \right],
\end{equation}
and applying the supersymmetry transformation $\us$ to this `new form' of the
action leads to
\begin{equation}
\begin{array}{l}
\us S^{\mathrm{lat}}_{\mathrm{BF}}
=\sum_\bx \left[ \epsilon_{\mu\nu}\Delta_{+\mu}\Delta_{+\nu}\uc(\bx)\uphi(\bx-\tilde{\ba}) \right. \\[10pt]
\,\,\,\,\,\, +\epsilon_{\mu\nu}\Delta_{+\mu}\uomega_{\nu}(\bx+\ba_{\nu}+2\ba))i\urho(\bx+\ba-\tilde{\ba})  \\[10pt]
\,\,\,\,\,\, +\ub(\bx+\ba)\Delta_{+\mu}\Delta_{-\mu}\uc(\bx) -\ub(\bx+\ba)\Delta_{+\mu}\Delta_{-\mu}\uc(\bx) \\[10pt]
\,\,\,\,\,\, \left. -i\urho(\bx+\ba-\tilde{\ba})\epsilon_{\rho\sigma}\Delta_{+\rho}\uomega_{\sigma}(\bx+ \ba_{\sigma}-\ba_1-\ba_2-\ba-\tilde{\ba}) \right] \\[10pt]
=  \sum_\bx i\urho(\bx+\ba-\tilde{\ba})\cdot\\[10pt]
\,\,\,\,\,\,\epsilon_{\mu\nu}\Delta_{+\mu}\left(\uomega_{\nu}(\bx+\ba_{\nu}+2\ba)-\uomega_{\nu}(\bx+\ba_{\nu})  \right) \\[10pt]
 \neq  0.
\end{array}
\end{equation}
Hence now the action appears not to be invariant
under the supersymmetry transformation. 

\subsubsection*{Some Observations}

We would like to add some interesting observations. For the mass terms in 
supersymmetric quantum mechanics the two
`good' representations $M$ and $\tilde{N}$, which are formally supersymmetry 
invariant,
come from the superfield formulation without (anti)commuting fields on the way. This seems
to be the general rule also for the interaction terms (then the second 
and third
line of (\ref{eq_action_lattice+D}) have to be modified accordingly). We
conjecture that for all the terms that appeared in the noncommutativity 
approach
so far (above and in the literature), there exists a `good' 
representation which
indeed is formally supersymmetry invariant.

After commuting fields one arrives at a `bad' representation,
like $\tilde{M}$, which is formally non-invariant.
However, when commuting fields in all terms of the sum
one can obtain a `good' representation again.
For instance, 
\begin{eqnarray}
\hat{M}=\phi(t)D(t)+i\psi_1(t)\psi_2(t+n)
\end{eqnarray}
which is obtained from the `good' $M$ by (anti)commuting the fields
in both terms, is invariant again.
We have seen an analogous effect in the interaction terms.

In the sum $M+\hat{M}$ all terms occur completely symmetrized
(taking into account minus signs when flipping two fermions).
This suggests to symmetrize every expression before applying
the supersymmetry transformations $s_A$ to it,
in order to be independent of starting with a `good' or `bad' representation
(e.g.\ symmetrizing the `bad' $\tilde{M}$ also leads to the `good' $\frac{1}{2}(M+\hat{M})$).
On the level of the expression itself the symmetrization $P$ is an identity operation, 
but
new supersymmetry transformations might be defined by $\bar{s}_A=s_AP$.
They do not suffer from the mentioned inconsistency, but will modify
$\Delta_\pm$ on the r.h.s.\ of the SUSY algebra and, moreover,
inherit another inconsistency from $P$, namely when acting on `composite fields'
they yield $\bar{s}_A[f_1(f_2f_3)]\neq\bar{s}_A[f_1f_2f_3]$.
Hence these formal manipulations do not remove the ambiguity in the
supersymmetry transformations, but one might view them as indications that the
noncommutativity approach will work in a more rigorous formulation.

%%%%%%%%%%%%%%%%%%%%%%%%%%%%%%%%%%%%%%%%%%%%%%%%%%%%%%%%%%%%%%%%%%

\section{Discussion and Conclusions}

We have presented the noncommutativity formalism to put
supersymmetric quantum mechanics on the lattice, 
including masses and typical field-theoretic
interactions. 
In the corresponding lattice action, Eq.\ (\ref{eq_action_lattice+D}),
the different powers of the
bosonic field $\phi$ come in very special combinations, see Eq.s 
(\ref{phia1})-(\ref{phia3}). 

At first sight the approach keeps all supersymmetries exact.
Indeed, showing the invariance of the lattice action in terms of superfields
(the analogue of Eq.\ \ref{eq_action_cont_super}) works as in the
continuum (using the Leibniz rule of $\epsilon^AQ_A$).
However, we have pointed to an inconsistency,
namely that the supersymmetry transformation of
products of component fields is not uniquely defined.
Therefore, the supersymmetry invariance of the lattice action in terms of
component fields is ambiguous.
We have shown this problem to be present in
the noncommutativity approach in general and also demonstrated it
extensively for (the mass term of) SUSYQM and the abelian super BF model.
We have not considered the recent SUSY lattice gauge theories \cite{DaddaGauge}, 
where there is a chance that the inconsistency is absent, because the link nature of the
variables prevents one from naively interchanging fields.

In our view the inconsistency is a remainder of the noncommutative algebra
$[t,\epsilon]\neq 0$. This structure is somewhat hidden in the
action, which seems to contain only functions of ordinary numbers (integers) $t$
as is very useful for numerical simulations. However, $t$ cannot be just a
number, as any number commutes with everything in any algebra.
It seems to be the too naive treatment of (fields as functions of) the coordinates which spoils the
definition of supersymmetry in such systems.

A natural step to establish a more profound formalism would be to promote all
objects to $T\times T$-matrices ($T$ being the number of lattice sites), to
replace the sum over $t$ by the trace and so on.
We have tried to represent the component fields by diagonal matrices and
$\epsilon$ (and $\theta$) by an off-diagonal one, in order to realize the
noncommutativity. However, the supersymmetry variations turned out to be either
trivially vanishing or to suffer from the same inconsistency.
The only way out seems to be non-sparse matrices, which increases the 
number of degrees of
freedom considerably (and slows down numerical simulations).

An alternative route to follow could be to define a modified product of fields,
as is typical for functions over noncommutative spaces.

\section*{Acknowledgements}

We are grateful to Alessandro d'Adda, Ernst-Michael Ilgenfritz, Pierre van Baal,
Peter Bongaarts, D\'{a}niel N\'{o}gr\'{a}di, Jan-Willem van Holten, Jan Smit, Simon Catterall, 
Ana Ach\'{u}carro and Alessandra Feo for useful discussions. 
FB has been supported by DFG (No. BR 2872/2-1) and MK by FOM.

{}

\end{document}